\documentclass[preprint,showpacs,preprintnumbers,amsmath,amssymb]{revtex4}
\usepackage{graphicx}
\usepackage{dcolumn}
\usepackage{bm}

\font\scripti=cmmi7
\font\scriptscripti=cmmi5
\def\sib#1{\setbox0 = \hbox{\scripti #1}
  \kern-.02em\copy0\kern-\wd0
  \kern.04em\box0} 
\def\ssib#1{\setbox0 = \hbox{\scriptscripti #1}
  \kern-.02em\copy0\kern-\wd0
  \kern.04em\box0} 
\font\tenib=cmmib10 
\skewchar\tenib='177 \skewchar\tenib='177 \skewchar\tenib='177
\def\pbold#1{\setbox0 = \hbox{$ #1 $}
  \kern-.022em\copy0\kern-\wd0
  \kern.011em\copy0\kern-\wd0
  \kern.011em\copy0\kern-\wd0
  \kern.011em\copy0\kern-\wd0
  \kern.011em\box0} 

\usepackage{graphicx}
\usepackage{dcolumn}
\usepackage{bm}

\def\ev#1{\langle #1 \rangle}
\def\cp{\mu}

\def\cpdwn{\mu_\downarrow}
\def\op{\Delta}
\def\s{\sigma}
\def\up{\uparrow}
\def\dwn{\downarrow}

\def\lesssim{\ \raise.3ex\hbox{$<$}\kern-0.8em\lower.7ex\hbox{$\sim$}\ }
\def\gesim{\ \raise.3ex\hbox{$>$}\kern-0.8em\lower.7ex\hbox{$\sim$}\ }

\begin{document}

\title{$\pi$-junction and spontaneous current state in a superfluid Fermi gas}

\author{Takashi Kashimura$^{1}$, Shunji Tsuchiya$^{2,3}$, and Yoji Ohashi$^{1,3}$}\affiliation{
$^1$Department of Physics, Keio University, 3-14-1 Hiyoshi, Kohoku-ku, Yokohama 223-8522, Japan,\\
$^2$Department of Physics, Tokyo University of Science, 1-3 Kagurazaka, Shinjuku-ku, Tokyo 162-8601, Japan,\\
$^3$CREST(JST), 4-1-8 Honcho, Saitama 332-0012, Japan} 

\date{\today}

\begin{abstract}

We discuss an idea to realize a spontaneous current in a superfluid Fermi gas. When a polarized Fermi superfluid ($N_\up > N_\dwn$, where $N_\sigma$ is the number of atoms in the hyperfine state described by pseudospin $\sigma=\uparrow, \downarrow$.) is loaded onto a ring-shaped trap with a weak potential barrier, some of excess atoms ($\Delta N=N_\uparrow-N_\downarrow$) are localized around the barrier. As shown in our previous paper [T. Kashimura, S. Tsuchiya, and Y. Ohashi, Phys. Rev. A \textbf{82}, 033617 (2010)], this polarized potential barrier works as a $\pi$-junction in the sense that the superfluid order parameter changes its sign across the barrier. Because of this, the phase of the superfluid order parameter outside the junction is shown to be twisted by $\pi$ along the ring, which naturally leads to a circulating supercurrent. While the ordinary supercurrent state is obtained as a metastable state, this spontaneous current state is shown to be more stable than the case with no current. Our results indicate that localized excess atoms would be useful for the manipulation of the superfluid order parameter in cold Fermi gases.
\end{abstract}

\maketitle

\section{Introduction} \label{sec1}
\begin{figure}[t]   
\begin{center}
\includegraphics[keepaspectratio,scale=0.7]{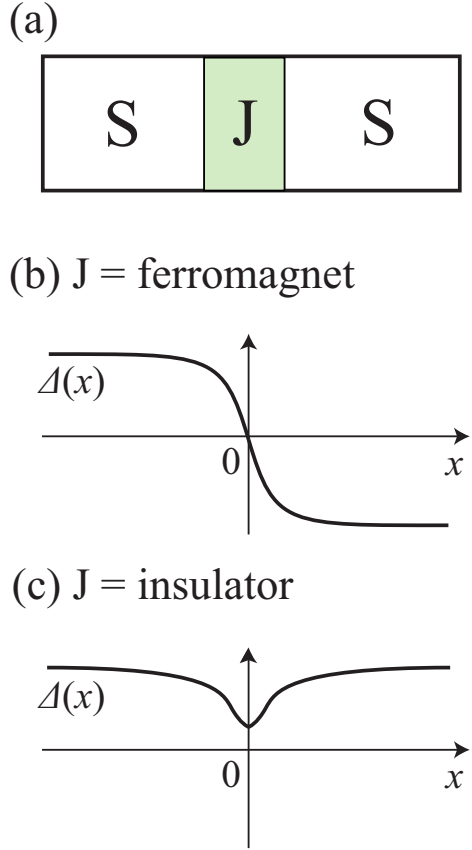}
\caption{(a) Schematic picture of Josephson junction, where `S' and `J' mean superconductor and Josephson junction, respectively. (b) When the junction (J) is made of ferromagnet (superconductor-ferromagnet-superconductor (SFS) junction), the $\pi$-junction may be realized, where the superconducting order parameter $\Delta(x)$ changes its sign across the junction ($\pi$-junction state). (c) When the ferromagnet is replaced by an insulator (superconductor-insulator-superconductor (SIS) junction), the sign change of $\Delta(x)$ does not occur ($0$-junction state).}
\label{fig1}
\end{center}    
\end{figure}

The $\pi$-junction \cite{Soda} is a typical superconducting device which uses magnetic effects on superconductivity. This is a superconductor-ferromagnet-superconductor junction \cite{Buzdin1,Bulaevskii,Buzdin2,Kanegae,Ryazanov,Oboznov}, as schematically shown in Fig. \ref{fig1}(a), which is different from the ordinary Josephson junction, where an insulator is used at the junction. The superconducting order parameter changes its sign across the $\pi$-junction, as shown in Fig. \ref{fig1}(b) \cite{note00}. This phenomenon is deeply related to the realization of the Fulde-Ferrell-Larkin-Ovchinnikov (FFLO) state \cite{FF,LO,Takada} inside the ferromagnetic junction. When the spatially oscillating FFLO order parameter takes opposite signs at the left and right edges of the ferromagnetic junction, the order parameters in the left and right superconductors also take opposite signs to each other. Because of this mechanism, the stability of $\pi$-junction state depends on the thickness of the ferromagnetic junction, which has been experimentally observed in a Nb-Cu$_{0.47}$Ni$_{0.53}$-Nb junction \cite{Oboznov}.
\par
\begin{figure}[t]   
\begin{center}
\includegraphics[keepaspectratio,scale=0.7]{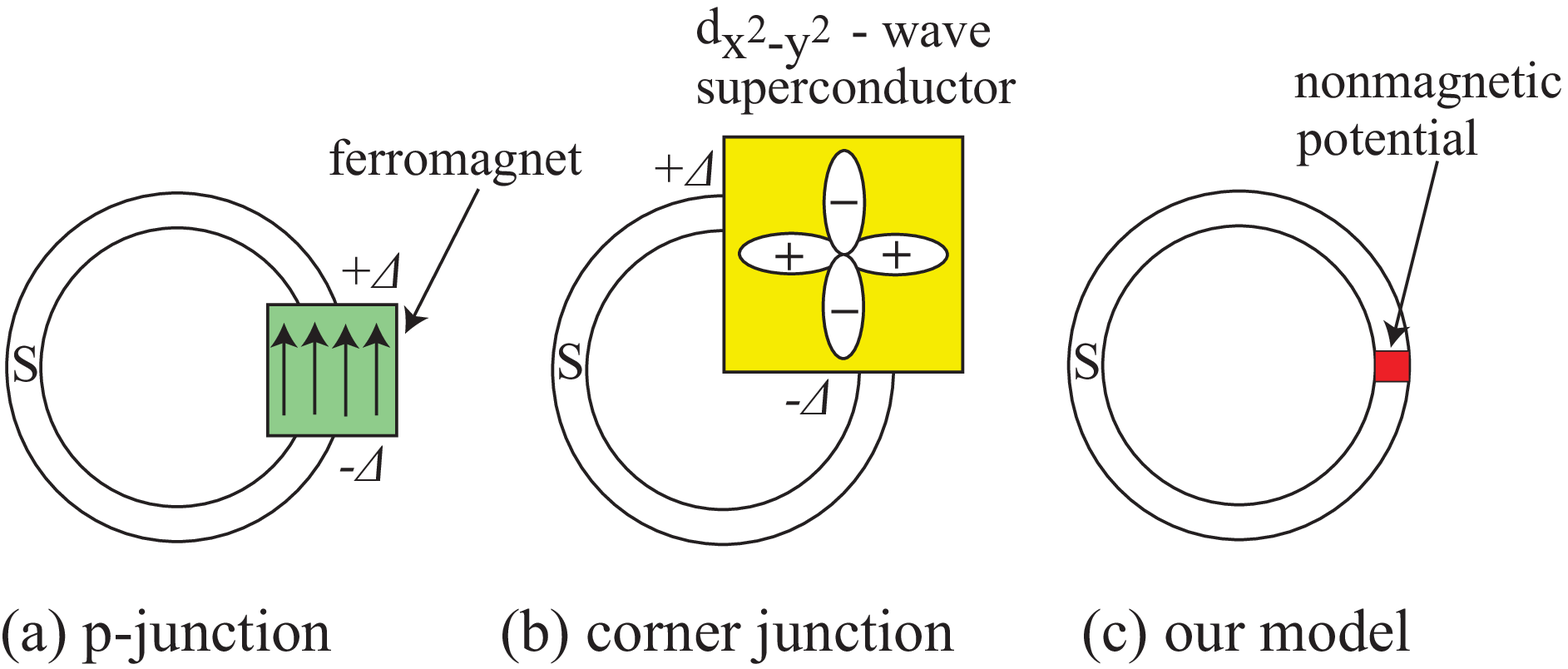}
\caption{Various systems that may realize spontaneous current states ($\pi$-ring state). (a) Ferromagnetic junction embedded in a superconducting ring. In this case, the sign change of the superconducting order parameter $\Delta$ across the ferromagnetic junction twists the phase of $\Delta$ along the ring, leading to a finite spontaneous circulating current. (b) Corner junction. The superconducting ring is connected to the (100) and (010) surfaces of high-$T_{\rm c}$ cuprates, and the sign change of the $d_{x^2-y^2}$-wave superconducting order parameter in momentum space leads to the phase twist. (c) Model superfluid Fermi gas in a ring-shaped torus trap with a weak nonmagnetic potential barrier. When the population imbalance exists, some excess atoms are localized around the barrier, which behave like a ferromagnetic junction.}
\label{fig2}
\end{center}    
\end{figure}

An interesting application of the $\pi$-junction is the induction of a spontaneous current \cite{Buzdin1}. When a $\pi$-junction is embedded in a superconducting ring as shown in Fig. \ref{fig2}(a), the $\pi$-junction twists the phase of the order parameter along the ring by $\pi$. This spatial phase modulation leads to a circulating Josephson current. Since the phase change along the ring equals $\pi$, a half-quantized vortex is trapped inside the ring. We note that this phenomenon is quite different from the case of SIS-junction. In this case, since the phase twist is absent [See Fig. \ref{fig1}(c).], the ordinary flux quantization $\phi_0=h/(2e)$ is obtained. While the supercurrent usually appears as a metastable state, the spontaneous current state is realized as a stable state. In this sense, the half-quantized vortex in the latter state is sometimes referred to as a self-induced vortex.
\par

We note that the half-quantized self-induced vortex associated with a spontaneous current has been experimentally observed \cite{Tsuei2} in the so-called corner-junction [Fig. \ref{fig2}(b)] \cite{Sigrist,Kato,Tsuei,Harlingen}. To twist the phase of the order parameter in the superconducting ring, the corner junction uses the property that the $d_{x^2-y^2}$-wave order parameter changes its sign in momentum space. We briefly note that the phase modulation by the corner-junction has played a crucial role in determining the pairing symmetry of high-$T_{\rm c}$ cuprates \cite{Sigrist,Kato,Tsuei,Harlingen}. 
\par
In this paper, we theoretically discuss an idea to realize the spontaneous current state ($\pi$-ring state) in a superfluid Fermi gas. Recently, this problem has been also discussed in a superfluid-(normal Fermi gas)-superfluid (SNS) junction within the framework of the Ginzburg-Landau theory \cite{Kulic}, under the assumption that the normal Fermi gas is polarized. In contrast to this previous work, in this paper, we do {\it not} assume a polarized junction from the beginning. Instead, we use the phase separation phenomenon observed in superfluid Fermi gases with population imbalance \cite{Zwierlein1,Partridge} ($N_\uparrow>N_\downarrow$, where $N_\sigma$ is the number of atoms in the hyperfine state described by pseudospin $\sigma = \uparrow, \downarrow$). As shown in our previous paper \cite{Kashimura}, when a {\it nonmagnetic} potential barrier is put in a superfluid Fermi gas with population imbalance, it is ``magnetized" in the sense that some of excess $\uparrow$-spin atoms are localized around it. Although the spin in this system is not a real spin, but a pseudospin describing an atomic hyperfine state, we clarified that the localized excess atoms behave like a ferromagnetic junction to cause the sign change of the superfluid order parameter across the potential barrier. Thus, when this pseudo-ferromagnetic junction can be realized in a ring-shaped trap, as schematically shown in Fig. \ref{fig2}(c)  \cite{Ryu}, one can expect a spontaneous circulating superflow. However, since the phase twist of the superfluid order parameter generally raises the gradient energy of the system, it is unclear whether the localization of excess atoms is energetically stable even when the phase twist occurs along the ring.
\par
To confirm our idea in a simple manner, we treat a two-component Fermi gas described by a one-dimensional ring Hubbard model with a weak nonmagnetic potential barrier. Within the mean-field theory at $T=0$, we self-consistently determine the superfluid order parameter, as well as the particle densities, for a given population imbalance ($N_\uparrow$,$N_\downarrow$). To examine the stability of the $\pi$-ring state, we energetically compare this state with the $0$-junction state (in which the sign change of the order parameter does not occur at the junction).
\par
Although the simple one-dimensional lattice model examined in this paper is actually different from real continuum Fermi gases, we emphasize that this simplification is not essential for our idea. However, to avoid lattice effects, we treat the case with low particle density. To examine effects of the dimensionality of the system, we also briefly examine a two-dimensional system.
\par
This paper is organized as follows. In Sec. II, we explain our formulation to examine the possibility of spontaneous current state in a superfluid Fermi gas loaded onto a one-dimensional ring-shaped trap. In Sec. III. we energetically compare the $\pi$-ring state with the $0$-junction state, and confirm that the former state is really possible. In this section, we also briefly deal with a two-dimensional system. Throughout this paper, we take $\hbar=1$, and the lattice constant to be unity, for simplicity.
\par

\section{Formulation} \label{sec2}
We consider a two-component Fermi gas described by the one-dimensional ring Hubbard Hamiltonian, 
\begin{equation} 
\begin{split}
H = -t \sum_{\ev{i,j},\sigma}
\left[
\hat{c}_{i,\sigma}^\dagger \hat{c}_{j,\sigma} + {\rm h.c.}
\right] 
-U 
\sum_{i} \hat{n}_{i,\uparrow } \hat{n}_{i,\downarrow} 
+\sum_{i,\sigma}
\left[V_i-\mu_\sigma
\right]\hat{n}_{i,\sigma }.
\label{eq.H}
\end{split} 
\end{equation}
Here, $\hat{c}_{i,\sigma}^\dagger$ is the creation operator of a Fermi atom in the atomic hyperfine state described by pseudospin $\sigma$ ($=\uparrow,\downarrow$) at the $i$-th site. Because of the ring geometry, the periodic boundary condition $c_{L_x+1,\sigma}=c_{1,\sigma}$ is imposed, where $L_x$ is the number of lattice site. $-t$ is a nearest-neighbor hopping and the summation $\ev{i,j}$ is taken over the nearest-neighbor pairs. $-U(<0)$ is an on-site attractive interaction. $\hat{n}_{i,\sigma}=\hat{c}_{i,\sigma}^\dagger \hat{c}_{i,\sigma}$ is the number operator at the $i$-th site. Since we consider a polarized Fermi gas, the chemical potential $\mu_\sigma$ depends on pseudospin $\sigma$. The {\it nonmagnetic} potential barrier is assumed to have the form,
\begin{equation} 
V_i=
V_0 e^{-(i-L_x/2)^2/\ell^2}.
\label{eq.pot}
\end{equation} 
We briefly note that the detailed spatial variation of Eq. (\ref{eq.pot}) is not crucial for our results\cite{Kashimura}.
\par
We treat the Hubbard model in Eq. (\ref{eq.H}) within the real-space mean-field theory. In our previous paper \cite{Kashimura}, we have explained the details of this mean-field theory, so that we only explain the outline here. For more details, we refer to Ref. \cite{Kashimura}. Introducing the superfluid order parameter $\Delta_i = U \ev{\hat{c}_{i,\dwn} \hat{c}_{i,\up}}$, as well as the particle densities $\ev{{\hat n}_{i,\sigma}}$ ($\sigma=\uparrow,\downarrow$), we obtain the mean-field Hamiltonian for Eq. (\ref{eq.H}) as 
\begin{eqnarray}
H_\text{MF} 
= 
&-&t \sum_{\langle i,j \rangle,\sigma} 
\left[
\hat{c}^\dagger_{i,\sigma} \hat{c}_{j,\sigma}+ {\rm h.c.} 
\right] 
-\sum_i
\left[
\Delta_i\hat{c}_{i,\up}^\dagger \hat{c}_{i,\dwn}^\dagger 
+ \Delta_i^*\hat{c}_{i,\dwn} \hat{c}_{i,\up}
\right]
\nonumber
\\
&+&\sum_{i,\sigma} 
\left[
V_i-\mu_\sigma-U \ev{\hat{n}_{i,-\s}} 
\right]
\hat{n}_{i,\s}
+ \sum_{i} 
\left[
\frac{|\op_i|^2}{U} + U \ev{\hat{n}_{i,\up}} 
\ev{\hat{n}_{i,\dwn}} 
\right].
\label{eq.MFH}
\end{eqnarray}
We diagonalize Eq. (\ref{eq.MFH}) by the real-space Bogoliubov transformation,
\begin{eqnarray}
\left(
\begin{array}{c}
{\hat c}_{1,\uparrow}\\
{\hat c}_{2,\uparrow}\\
\vdots\\
{\hat c}_{M,\uparrow}\\
{\hat c}_{1,\downarrow}^\dagger\\
{\hat c}_{2,\downarrow}^\dagger\\
\vdots\\
{\hat c}_{M,\downarrow}^\dagger\\
\end{array}
\right)
=
{\hat W}
\left(
\begin{array}{c}
{\hat \alpha}_{1,\uparrow}\\
{\hat \alpha}_{2,\uparrow}\\
\vdots\\
{\hat \alpha}_{M,\uparrow}\\
{\hat \alpha}_{1,\downarrow}^\dagger\\
{\hat \alpha}_{2,\downarrow}^\dagger\\
\vdots\\
{\hat \alpha}_{M,\downarrow}^\dagger\\
\end{array}
\right).
\label{eq.BB}
\end{eqnarray}
Here, $\hat{W}$ is a $2M \times 2M$ unitary matrix, where $M$ is the total number of lattice sites. In the present one-dimensional case, one finds $M=L_x$. After the diagonalization, we have
\begin{eqnarray}
H_\text{MF}
&=&
\sum_{j,\sigma}E_{j,\s} \hat{\alpha}_{j,\s}^\dagger \hat{\alpha}_{j,\s} 
\nonumber
\\
&+& \sum_{i=1}^M 
\bigg[
V_i-\cpdwn - U\ev{\hat{n}_{i,\up}} + \frac{|\op_i|^2}{U} 
+ U \ev{\hat{n}_{i,\up}} \ev{\hat{n}_{i,\dwn}} 
 -E_{i,\downarrow}
\bigg],
\label{DiagonalH}
\end{eqnarray}
where $E_{j,\sigma}$ is the eigen-energy of the $j$-th Bogoliubov excitation. Using the unitary matrix ${\hat W}$, one calculates the superfluid order parameter $\Delta_i$, as well as the particle density $\ev{\hat{n}_{i,\sigma}}$, as
\begin{eqnarray}
\Delta_i 
&=& U \sum_{j=1}^M
\Bigl[
W_{i,j} W_{M+i,j}^* \Theta( -E_{j,\up} ) 
+ W_{i,M+j} W_{M+i,M+j}^* \Theta(E_{j,\downarrow})
\Bigr],
\label{GAP}
\\
\ev{\hat{n}_{i,\up}} &=& 
\sum_{j=1}^M 
\Bigl[
|W_{i,j}|^2 \Theta( -E_{j,\up} )+ |W_{i,M+j}|^2 \Theta(E_{j,\downarrow})
\Bigr], 
\label{n_up}
\\
\ev{\hat{n}_{i,\dwn}} &=&
\sum_{j=1}^M 
\Bigl[
|W_{M+i,j}|^2 \Theta( E_{j,\up} ) 
+ |W_{M+i,M+j}|^2 \Theta(-E_{j,\downarrow}) 
\Bigr],
\label{eq.n_down}
\end{eqnarray}
where $\Theta(E)$ is the step function. The total number $N_\sigma$ of $\sigma$-spin atoms is then given by
\begin{equation}
N_\sigma=\sum_{i}\ev{\hat{n}_{i,\sigma}}.
\label{eq.num}
\end{equation}
\par
For a given parameter set of $(U,N_\uparrow,N_\downarrow,V_0, \ell)$, we numerically diagonalize Eq. (\ref{eq.MFH}), and self-consistently determine $\Delta_i$, $\ev{\hat{n}_{i,\up}}$, and $\ev{\hat{n}_{i,\dwn}}$ by using Eqs.  (\ref{GAP})-(\ref{eq.n_down}). To examine the stability of spontaneous current state, we calculate the energy \cite{Kashimura},
\begin{equation} \label{Legendre} \begin{split}
E_G
&=
\sum_{j,\sigma}E_{j,\sigma} \Theta (-E_{j,\sigma})
+ \sum_{\s} \cp_\s N_\s
\\
&+ \sum_{i} 
\Big[ V_i - \mu_\downarrow - U \langle \hat{n}_{i,\uparrow } \rangle
+ {|\Delta_i|^2 \over U} 
+ U \langle \hat{n}_{i,\uparrow } \rangle 
\langle \hat{n}_{i,\downarrow} \rangle 
 - E_{i,\downarrow}
\Big].
\end{split} \end{equation}
\par
Once a self-consistent solution is obtained, the current density $J_i$ flowing in the ring is given by \cite{Mahan}
\begin{eqnarray}
J_i 
&=&
it \sum_\sigma \left[ \ev{\hat{c}_{i+1,\sigma}^\dagger \hat{c}_{i,\sigma}} - \ev{\hat{c}_{i,\sigma}^\dagger \hat{c}_{i+1,\sigma}} \right] 
\nonumber
\\
&=& -2t \sum_j 
\text{Im} 
\Bigl[ W^*_{i+1,j} W_{i,j} \Theta (-E_{j,\up})
+W^*_{i+1,M+j} W_{i,M+j} \Theta (E_{j,\dwn})
\nonumber
\\
&{}&~~~~~~~~~~~~+W_{M+i+1,j} W_{M+i,j}^* \Theta (E_{j,\up})
+W_{M+i+1,M+j} W_{M+i,M+j}^* \Theta (-E_{j,\dwn})
\Bigr].
\end{eqnarray}
\par
\begin{figure}[t]   
\begin{center}
\includegraphics[keepaspectratio,scale=0.6]{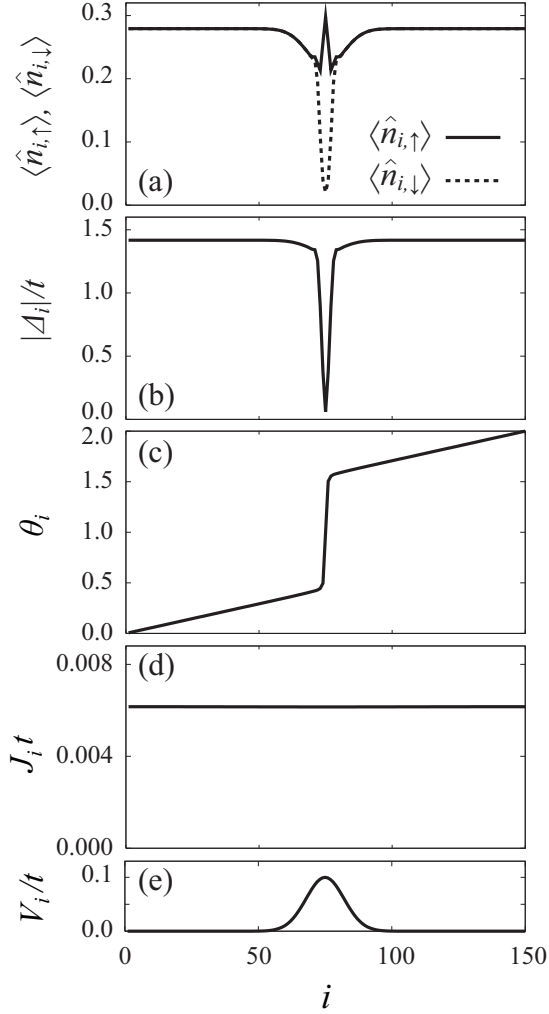}
\caption{Calculated stable spontaneous current state in a spin-polarized superfluid Fermi gas loaded onto a one-dimensional ring trap. (a) Particle density $\langle n_{i,\sigma}\rangle$. (b) Magnitude of superfluid order parameter $|\Delta_i|$. (c) Phase $\theta_i$ of the order parameter $\Delta_i=|\Delta_i|e^{i\theta_i}$. (d) Supercurrent density $J_i$. (e) Spatial variation of the nonmagnetic potential barrier $V_i$. The polarization is taken as $(N_\uparrow,N_\downarrow)=(41,40)$. In this case, one excess $\uparrow$-spin atom is localized around the potential barrier. For the other parameters, we set $L_x=150$, $U/t=4$, $V_0/t=0.1$, and $\ell=10$. These values are also used in Fig.\ref{fig4}.}
\label{fig3}
\end{center}    
\end{figure}

\begin{figure}[t]   
\begin{center}
\includegraphics[keepaspectratio,scale=0.6]{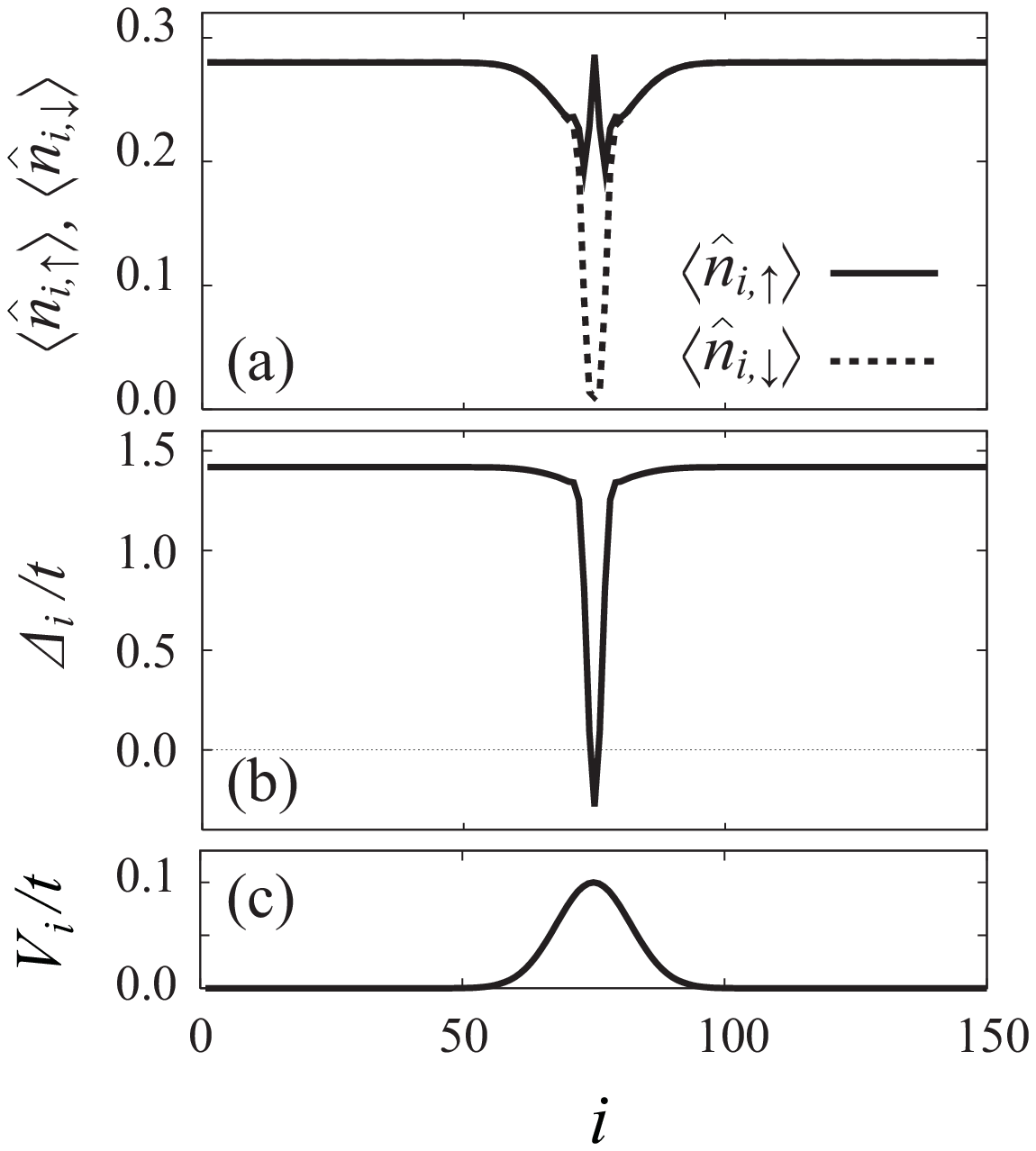}
\caption{Solution for the $0$-junction state. (a) Particle density $\ev{\hat{n}_{i,\sigma}}$. (b) Superfluid order parameter $\Delta_i$. (c) Potential barrier $V_i$. In this case, the phase twist is absent in the ring, so that one obtains $J_i=0$.}
\label{fig4}
\end{center}    
\end{figure}

\section{$\pi$-ring state and spontaneous current} \label{sec3}
\par
Figure \ref{fig3} shows the calculated spontaneous current state ($\pi$-ring state) when $N_\uparrow=41>N_\downarrow=40$. In panel (a), the excess $\uparrow$-spin atom ($N_\uparrow-N_\downarrow=1$) is found to be localized around the barrier at $i=75$. Because of this ``local magnetization", the superfluid order parameter is remarkably suppressed around it, as shown in panel (b). When we simply call the spatial region where the magnetization is large the ferromagnetic junction, panel (c) indicates that the phase $\theta_i$ of the order parameter approximately changes by $\pi$ across the junction. That is, the $\pi$-junction is realized. To satisfy the single-valueness of $\Delta_i$, the phase $\theta_i$ also varies outside the ferromagnetic junction, as shown in panel (c). Because of this phase modulation,  panel (d) shows that a spontaneous current flows. In this panel, the induced current is found to be uniform, although the spatial variation of $\theta_i$ is not uniform, especially around the junction. This means that the current is conserved in the ring. 
\par
In addition to the $\pi$-ring state in Fig. \ref{fig3}, we also obtain the $0$-junction state shown in Fig. \ref{fig4}. In the latter state, although the excess $\uparrow$-spin atom is also localized around the barrier, it does not induce the sign change of the superfluid order parameter across the ferromagnetic junction. As a result, the supercurrent $J_i$ is absent in this case. Comparing the energy $E_G^\pi$ of the $\pi$-junction state with the energy $E_G^0$ of the $0$-junction state, we obtain
\begin{equation}
\Delta E_G=E_G^\pi-E_G^0=-0.013t<0.
\label{eq.0-pi}
\end{equation}
Thus, in contrast to the ordinary {\it metastable} supercurrent state, the $\pi$-ring state with a spontaneous superflow is more stable than the $0$-junction state with no current.
\par
\begin{figure}[t]   
\begin{center}
\includegraphics[keepaspectratio,scale=0.6]{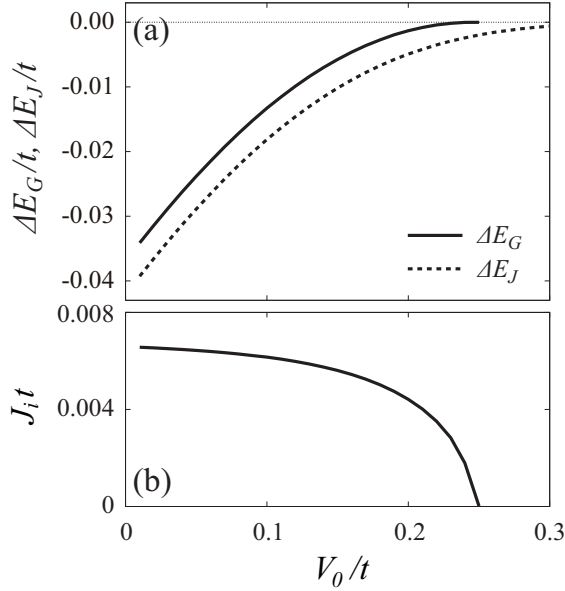}
\caption{(a) Energy difference between the $\pi$-ring state and 0-junction state, $\Delta E_G \equiv E_G^{\pi} - E_G^\text{0}$, as a function of the barrier height $V_0$ (solid line). The dashed line shows $\Delta E_J$ in Eq. (\ref{eq.ej}), describing the difference of junction energies between the $\pi$-junction and $0$-junction in the absence of the phase twist. This quantity is evaluated by considering the model shown in Fig. \ref{fig8}(b). (b) Spontaneous current density $J_i$ in the $\pi$-ring state, as a function of the potential height $V_0$. We take $L_x=150$, $U/t=4$, and $\ell=10$. These parameters are also used in Figs. \ref{fig6} and \ref{fig7}.}
\label{fig5}
\end{center}    
\end{figure}

\begin{figure}[t]   
\begin{center}
\includegraphics[keepaspectratio,scale=0.65]{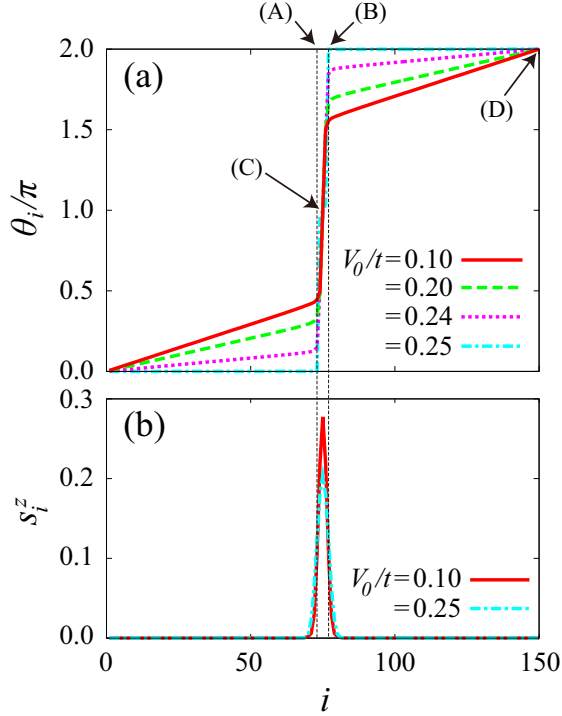}
\caption{(a) Spatial variation of the phase $\theta_i$ of the superfluid order parameter $\Delta_i=|\Delta_i|e^{i\theta_i}$ in the $\pi$-ring state. (b) Local magnetization $s^z_i=\langle n_{i,\uparrow}\rangle-\langle n_{i,\downarrow}\rangle$. The dashed lines (A) and (B) show the positions at which the polarization equals half the maximum value of $s^z_i$. Although there is no clear boundary between the magnetized junction with $s^z_i\ne 0$ and the superfluid ring with $s^z_i=0$ in the present case, we simply call the region between the lines (A) and (B) the ferromagnetic junction.
}
\label{fig6}
\end{center}    
\end{figure}

\begin{figure}[t]   
\begin{center}
\includegraphics[keepaspectratio,scale=0.65]{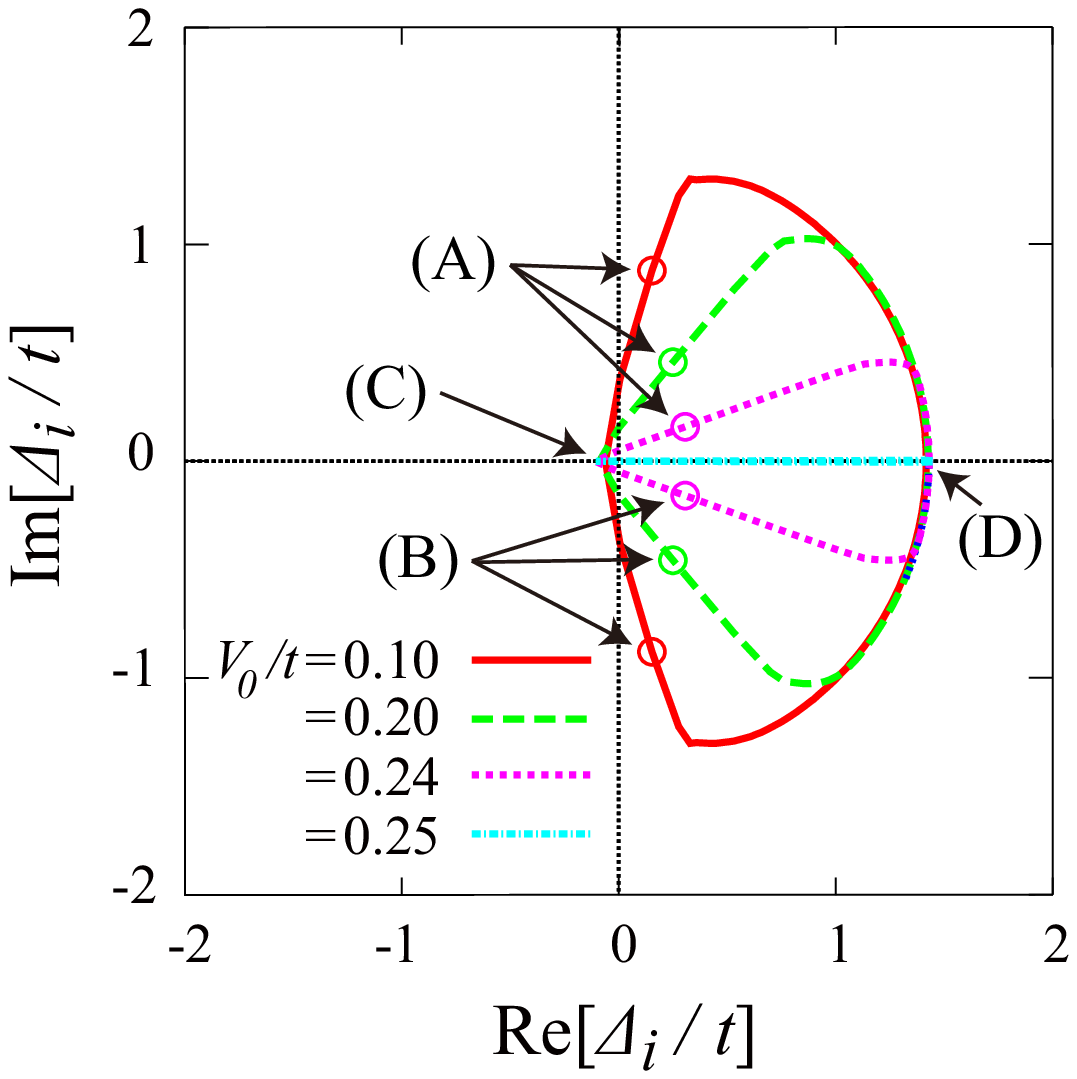}
\caption{Locus of the superfluid order parameter $\Delta_i$ in the ${\rm Re}[\Delta_i]$-${\rm Im}[\Delta_i]$-plane. The arrows with (A)-(D) show $\Delta_i$ at the positions shown in Fig. \ref{fig6}(a).
}
\label{fig7}
\end{center}    
\end{figure}
\par
Evaluating the total angular momentum $\langle L_z\rangle$ of the spontaneous current state shown in Fig. \ref{fig3}, we obtain 
\begin{equation}
\langle L_z\rangle \simeq mR J_T\simeq{RJ_T \over 2t}=11.0~~~(\hbar=1),
\label{eq.Lz}
\end{equation}
where $R=L_x/(2\pi)=150/(2\pi)$ is the radius of the ring, and $J_T=\sum_i J_i=L_xJ_i=150\times(6.16t\times 10^{-3})$ is the total current, which is given by summing up the local current density $J_i$ over the ring. Since we consider the case of low particle density, the particle mass $m$ is approximated to $m\simeq (2t)^{-1}$ in Eq. (\ref{eq.Lz}). At a glance, this finite angular momentum looks contradicting the so-called Bohm's theorem \cite{Bohm,Ohashi1,note10}, stating the vanishing total angular momentum in thermodynamically stable states. In this regard, we note that, while the Bohm's theorem assumes $J_T=O(R)$ for a fixed particle density \cite{Bohm,Ohashi1}, one finds $J_T=O(R^0)$ in the present case \cite{note11}. Including this difference, we find that, in the present case, the Bohm's theorem gives the upper limit of the total angular momentum $\langle L_z \rangle$ given by
\begin{equation}
\Bigl|
{2\langle L_z\rangle \over N}
\Bigr|<1,
\label{eq.LL}
\end{equation}
where $N$ is the total number of particles contributing to the superflow. (For the derivation of Eq. (\ref{eq.LL}), see the Appendix.) In the case of Fig. \ref{fig3} ($N=80$ \cite{note12}), we obtain $|2\langle L_z\rangle/N|=0.275< 1$, which satisfies Eq. (\ref{eq.LL}). Thus, the present spontaneous current state does not contradict the Bohm's theorem.
\par
While the $\pi$-junction lowers the junction energy, the phase twist along the ring raises the gradient energy of the condensate. Thus, the stability of the spontaneous current state is determined as a result of the competition between these two energies. Indeed, when the strength of the Josephson coupling at the junction is decreased by increasing the barrier height $V_0$, Fig. \ref{fig5}(a) shows that the energy difference $\Delta E_G$ in Eq. (\ref{eq.0-pi}) becomes small. At the same time, the magnitude $J_i$ of the spontaneous flow also becomes small, as shown in Fig. \ref{fig5}(b). Within the accuracy of our numerical calculations, the finite spontaneous current is not obtained when $V_0/t\gesim 0.25$. 
\par
The vanishing supercurrent for $V_0/t\gesim 0.25$ can be also understood from the spatial variation of the phase $\theta_i$ of the superfluid order parameter shown in Fig. \ref{fig6}(a). In this figure, we find that the spatial variation of $\theta_i$ gradually becomes weak outside the magnetized junction [outside the region between (A) and (B)], as one increases the barrier height $V_0$. The phase $\theta_i$ eventually becomes constant as $\theta_i=0$ ($i\lesssim 75$) and $\theta_i=2\pi$ ($i\gesim 75$) when $V_0/t= 0.25$. Since the Josephson current originates from the phase gradient, the supercurrent no longer flows.
\par
We briefly note that, although the spontaneous current state looks stable in the low potential limit $V_0 \rightarrow 0$ in Fig. \ref{fig5}(a), this is peculiar to the present one-dimensional model. In the phase separation regime of a one-dimensional system with no potential barrier, once excess atoms are localized somewhere in the system, it immediately works as a magnetic junction, leading to the $\pi$-ring state. However, in the case of higher dimensional systems, although localized excess atoms may behave like localized impurities, they do not form a magnetic  barrier in the absence of a wall potential, so that the $\pi$-junction and the $\pi$-ring state are not realized.
\par
As explained previously, the origin of the spontaneous current is the sign change of the superfluid order parameter across the $\pi$-junction. In this regard, we note that the phase change across the junction between (A) and (B) in Fig. \ref{fig6}(a) actually exceeds $\pi$, which becomes more remarkable for higher barrier height $V_0$. This means that, in addition that the phase $\theta_i$ outside the junction is affected by the $\pi$-junction, the $\pi$-junction is also affected by the attached ring. To see this more clearly, we show in Fig. \ref{fig7} the locus of the superfluid order parameter in the complex ${\rm Re}[\Delta_i]$-${\rm Im}[\Delta_i]$ plane. Moving from the left edge of the junction (A) to the right edge (B) through the junction center (C), one finds that the phase change is larger than $\pi$. The angle (A)-(C)-(B) reaches $2\pi$ when $V_0/t=0.25$, at which the spontaneous current vanishes, as shown in Fig. \ref{fig5}(b). 
\par

\begin{figure}[t]   
\begin{center}
\includegraphics[keepaspectratio,scale=0.65]{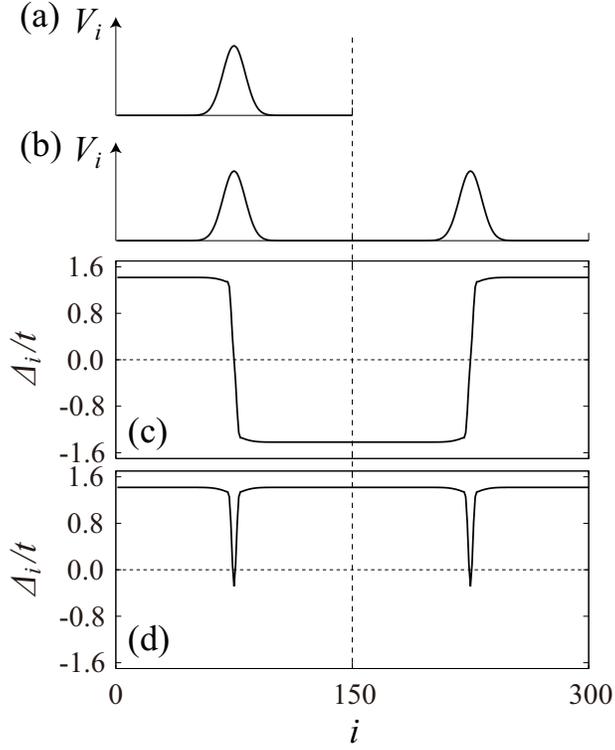}
\caption{(a) Model one-dimensional system used in calculating Fig. \ref{fig5}. (b) One-dimensional ring system to evaluate the junction energy. The system size $L_x=300$ is twice as long as the case of (a). In both (a) and (b), the periodic boundary condition is imposed. (c) Self-consistent solution with two $\pi$-junctions. (d) Self-consistent solution with two 0-junctions. In the present case, one finds $E_G^{\pi \times 2}<E_G^{0 \times 2}$. In panels (c) and (d), we take $(N_\uparrow, N_\downarrow)=(82, 80)$, so as to gives the same particle density as that in the cases of Figs. \ref{fig3} and \ref{fig4}.}
\label{fig8}
\end{center}    
\end{figure}
\par
To examine the competition between the energy gain at the $\pi$-junction and the gradient energy associated with the phase modulation along the ring, we evaluate the latter energy. For this purpose, it is convenient to treat the ring system which is twice as long as that examined in Fig. \ref{fig5} with two potential barriers, shown in Fig. \ref{fig8}(b). Setting the same particle density and the population imbalance as in the case of Fig. \ref{fig5}, we obtain the stable solution with two $\pi$-junctions and the energy $E_G^{\pi\times 2}$ (Fig.\ref{fig8}(c)), as well as the solution with two $0$-junctions and the energy $E_G^{0\times 2}$ [Fig. \ref{fig8}(d)]. In Fig. \ref{fig8}(c), since the sign change of $\Delta_i$ occurs twice, the phase modulation outside the magnetized junctions, as well as the associated spontaneous current, are absent. Thus, the difference between the quantity
\begin{equation}
\Delta E_J={1 \over 2}[E_G^{\pi \times 2}-E_G^{0 \times 2}]
\label{eq.ej}
\end{equation}
and $\Delta E_G$ shown in Fig. \ref{fig5}(a) is considered to give the gradient energy. Indeed, the difference between the solid line ($\Delta E_G$) and the dashed line ($\Delta E_J$) indicates that the phase twist increases the energy of the system. In particular, the fact that $\Delta E_J$ is negative even for $V_0/t\ge 0.25$ in Fig. \ref{fig5}(a) means that the gradient energy destroys the $\pi$-ring state there.
\par
\begin{figure}[t]   
\begin{center}
\includegraphics[keepaspectratio,scale=0.4]{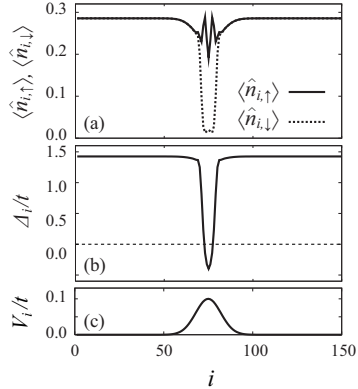}
\caption{Stable 0-junction state when $N_\uparrow=42>N_\downarrow=40$. For the other parameters, we use the same values as those used in Fig.\ref{fig3}. (a) Particle density $\langle n_{i,\sigma}\rangle$. (b) Superfluid order parameter $\Delta_i$. (c) Potential barrier $V_i$. Because of the thicker ferromagnetic junction than the case of Fig.\ref{fig3} ($N_\uparrow=41$, $N_\downarrow=40$), the 0-junction state becomes more stable than the $\pi$-junction state.}
\label{fig9}
\end{center}    
\end{figure}

\begin{figure}[t]   
\begin{center}
\includegraphics[keepaspectratio,scale=0.6]{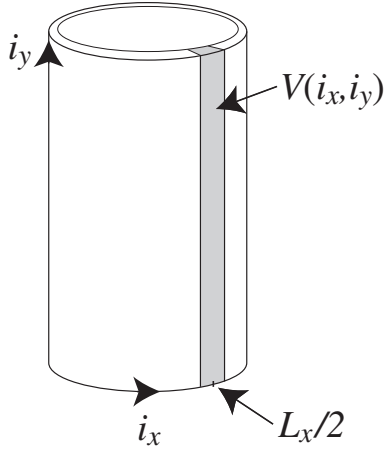}
\caption{Model two-dimensional cylinder-type lattice system. The number of lattice sites in the $y$-direction equals $L_y$, and the periodic boundary condition is imposed in this direction. $V(i_x,i_y)=V_0e^{-(i_x-L_x/2)^2/\ell^2}$ is a barrier potential, where $L_x$ is the number of lattice sites in the $x$-direction, and $\ell$ describes the width of the nonmagnetic potential barrier in the $x$-direction. $(i_x,i_y)$ describes the position of a lattice site in the square lattice. The barrier potential is uniform in the $y$-direction, and we consider a spontaneous current flowing in the $x$-direction.}
\label{fig10}
\end{center}    
\end{figure}

\begin{figure}[t]   
\begin{center}
\includegraphics[keepaspectratio,scale=0.6]{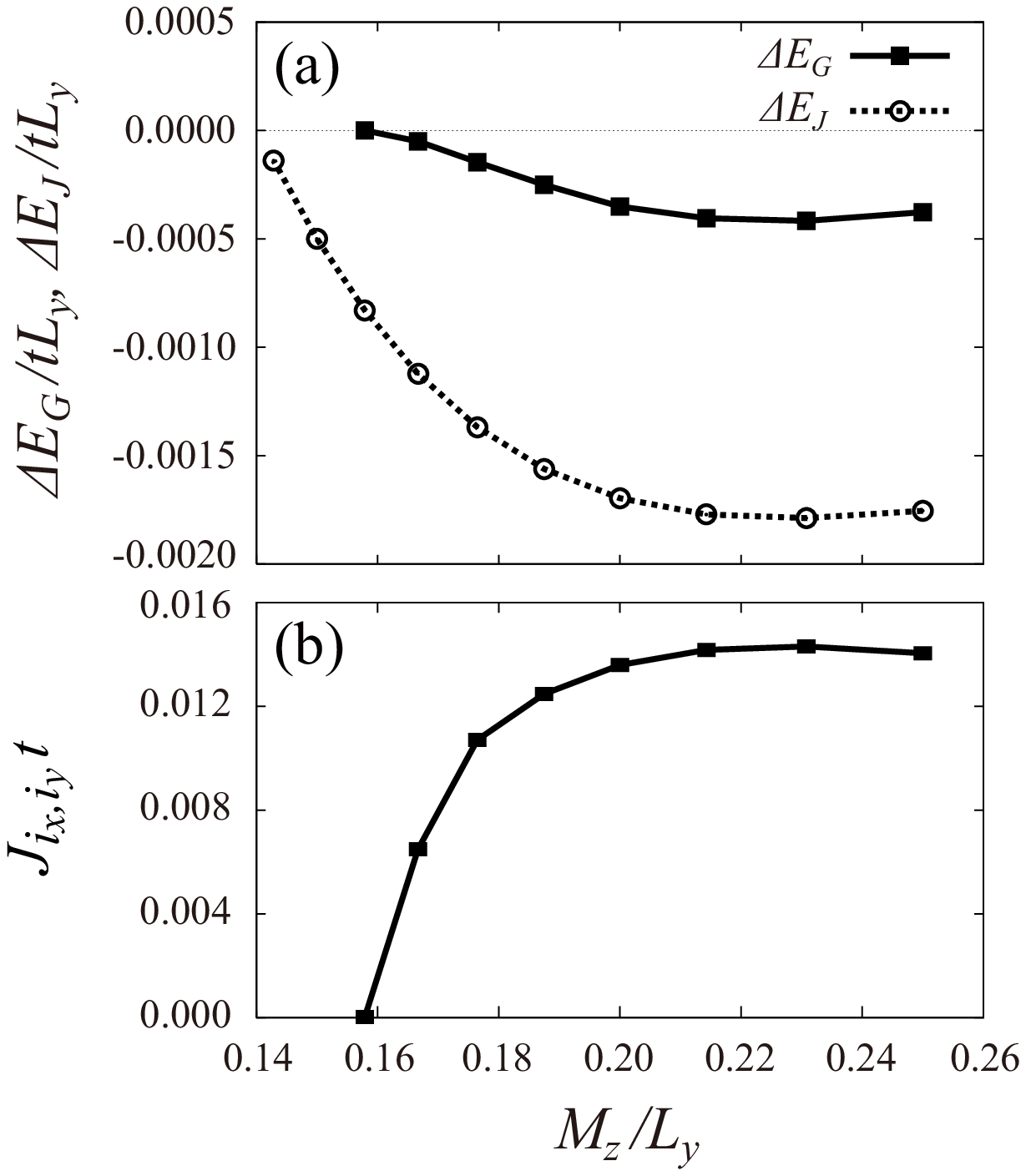}
\caption{(a) solid line: calculated energy difference $\Delta E_G$ between the $\pi$-ring state and the 0-junction state, as a function of the magnetization density $M_z/L_y$ per lattice site in the $y$-direction (where $M_z=N_\uparrow-N_\downarrow$). The dashed line shows $\Delta E_J$ in Eq. (\ref{eq.ej}). To evaluate $\Delta E_J$, we have used the method explained above Eq. (\ref{eq.ej}). (b) Spontaneous current density $J_{i_x,i_y}$ at the $(i_x, i_y)$ site, flowing in the $x$-direction. This quantity is calculated from $J_{i_x, i_y} 
=it \sum_\sigma \left[ \ev{\hat{c}_{i_x+1, i_y, \sigma}^\dagger \hat{c}_{i_x, i_y, \sigma}} - \ev{\hat{c}_{i_x, i_y, \sigma}^\dagger \hat{c}_{i_x+1, i_y, \sigma}} \right] $, where $\hat{c}_{i_x, i_y, \sigma}$ is the annihilation operator at the $(i_x, i_y)$ site. In this figure, we take $L_x=50$, $U/t=7$, $V_0/t=0.1$, and $\ell=2$. To tune the magnetization $M_z/L_y$ without changing the particle density far away from the junction, we set $N_\uparrow=N_\downarrow+3$ and $N_\downarrow=5L_y$, and vary the number of lattice site in the $y$-direction between $L_y=[12,19]$. 
}
\label{fig11}
\end{center}    
\end{figure}

\par
In addition to the potential height $V_0$, the magnitude of the magnetization at the junction is also expected to affect the stability of spontaneous current state. However, in examining this problem, the present one-dimensional model is not useful, because the thickness of the ferromagnetic junction in this model is known to remarkably depend on the population imbalance \cite{Kashimura}. Then, since the stability of $\pi$-junction is deeply related to the ratio of the junction thickness to the wavelength of the FFLO oscillation \cite{Buzdin1}, the so-called 0-$\pi$ transition frequently occurs, as one increases the population imbalance. Indeed, when we add one more $\uparrow$-spin atom to the system examined in Fig.\ref{fig3}, the magnetized junction becomes thicker, giving the stable $0$-junction state shown in Fig. \ref{fig9}.
\par
To avoid this problem peculiar to the present one-dimensional model, we consider a two-dimensional cylinder-type lattice system shown in Fig. \ref{fig10}. In this model, since the excess $\uparrow$-spin atoms localized around the barrier may align in the $y$-direction, we can safely tune the magnetization of the junction without the 0-$\pi$ transition. 
\par
Figure \ref{fig11} shows the stability of the spontaneous current state as a function of the magnetization density at the junction per lattice site in the $y$-direction, $M_z/L_y$, where $M_z=N_\uparrow-N_\downarrow$, and $L_y$ is the number of lattice sites in the $y$-direction. (Note that all the excess $\uparrow$-spin atoms are localized around the barrier in the present case.) As expected, while the stable spontaneous current state ($\Delta E_G<0$ and $J_{i_x, i_y}>0$, where $J_{i_x, i_y}$ is the current density at the $(i_x, i_y)$ site, flowing in the $x$-direction.) is obtained when the magnetization density is large to some extent ($M_z/L_y> 0.158$), such a $\pi$-ring solution is not obtained when $M_z/L_y\le 0.158$. Because of $\Delta E_J<0$ even for $M_z/L_y\le 0.158$, the vanishing spontaneous current in this regime is due to the gradient energy by the phase twist, as in the one-dimensional case. 
\par
So far, we have examined the case at $T=0$, in order to confirm our idea in a simple manner. In this regard, we note that the spontaneous current state is also possible at finite temperatures, when the $\pi$-junction is realized. Indeed, extending our theory to the finite temperature region, we find that, in the case of Fig. \ref{fig3}, the $\pi$-ring state is stable to $T \simeq 0.11 T_\text{c}$, where $T_\text{c}$ is the superfluid phase transition temperature evaluated in the uniform and unpolarized case with $N_\up=N_\dwn=40$ and $U/t=4$ (although we don't show the numerical result here). When $T\gesim 0.11T_{\rm c}$, the magnetization of the potential barrier vanishes, so that the $\pi$-junction state is no longer obtained. We briefly note that the vanishing excess atoms localized around the potential barrier is consistent with the phase diagram of a polarized superfluid Fermi gas \cite{Shin}, where the transition from the phase separation regime to a polarized superfluid regime occurs at a finite temperature, depending on the magnitude of the population imbalance.  
\par
\section{summary} \label{sec5}

To summarize, we have discussed the possibility of spontaneous current state in a superfluid Fermi gas. In a polarized Fermi superfluid loaded onto a ring trap with a weak nonmagnetic barrier, we showed that excess atoms localized around the barrier works as a ferromagnet to twist the phase of the order parameter by $\pi$. This phase modulation induces a spontaneous circulating supercurrent. In contrast to the ordinary supercurrent state (which is a metastable state), the present spontaneous current state is more stable than the state with no superflow. To confirm our idea in a simple manner, we have examined one- and two-dimensional ring Hubbard models within the mean-field theory at $T=0$. However, the dimensionality, as well as the presence of background lattice, are not essential for our results.
\par
The spontaneous supercurrent obtained in this paper is a direct consequence of the phase twist along the ring by the sign change of the superfluid order parameter at the $\pi$-junction, which is made of excess $\uparrow$-spin atoms. Thus, the observation of the former phenomenon would be also an evidence of the formation of the (pseudo)ferromagnetic junction working as a $\pi$-junction. In this regard, while the $\pi$-junction locally appears at the barrier, the phase modulation occurs over the entire system, so that the observation of the latter phenomenon would be easier. In particular, the phase modulation might be observed by using the interference of two condensates. As discussed in the introduction, since the $\pi$-junction is a typical device which uses competition between superconductivity and ferromagnet, our results would be useful for examining physical properties of this unique device using cold Fermi gases.

\section*{Acknowledgements}

We would like to thank S. Watabe, D. Inotani, and R. Watanabe for useful discussions. T. K. was supported by Global COE Program "High-Level Global Cooperation for Leading-Edge Platform on Access Space (C12)". Y. O. was supported by Grant-in-Aid for Scientific research from MEXT in Japan (22540412, 23104723).


\begin{figure}[t]   
\begin{center}
\includegraphics[keepaspectratio,scale=0.65]{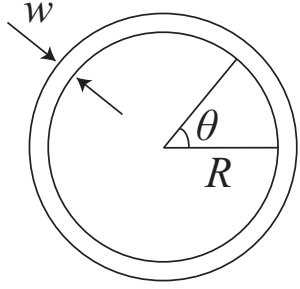}
\caption{Model ring system to explain the Bohm's theorem. We assume that the radius $R$ of the ring is much larger than the ring width $w$.}
\label{fig12}
\end{center}    
\end{figure}

\appendix
\section{Derivation of Eq. (\ref{eq.LL})} \label{app}
\par
To explain the outline of the proof of the Bohm's theorem, we consider a continuum Fermi gas in a ring shown in Fig. \ref{fig12}. The Hamiltonian is given by
\begin{equation} \begin{split}
H &= \sum_\sigma \int d \bm{r} \psi_\sigma^\dagger (\bm{r}) 
\left[ \frac{\hat{\bm{p}}}{2m} + V(\bm{r})  - \mu_\sigma \right] 
\psi_\sigma (\bm{r}) \\
&+ \frac{1}{2} \sum_{\sigma, \sigma'} \int d \bm{r} d \bm{r}' 
\psi_\sigma^\dagger (\bm{r}) \psi_\sigma^\dagger (\bm{r}') U(\bm{r}-\bm{r}')
\psi_\sigma (\bm{r}') \psi_\sigma (\bm{r}).  
\end{split} 
\label{eq.a1}
\end{equation}
Here, $\psi_\sigma(\bm{r})$ is the fermion field operator with pseudospin $\sigma$ $(=\uparrow,\downarrow)$, $\hat{\bm{p}}= -i \nabla$, and $m$ is an atomic mass. $V({\bm r})$ is a potential, and $U(\bm{r}-\bm{r}')$ is an interaction between atoms. We assume that the chemical potential $\mu_\sigma$ may depend on pseudospin $\sigma$ to include the case of population imbalance.
\par
Following Ref. \cite{Bohm}, we assume that the ground state $|G \rangle$ has a finite total angular momentum. Using this assumed ground state, we introduce the following trial state,
\begin{equation} \begin{split}
|T \rangle \equiv \exp{ \left[ in \sum_\sigma \int d \bm{r} 
\psi_\sigma^\dagger (\bm{r}) \theta \psi_\sigma (\bm{r}) \right] }  |G \rangle,
\label{eq.a2}
\end{split} \end{equation}
where $n$ is taken to be an integer so as to guarantee the single-valueness of $|T\rangle$. According to the variational principle, if $|G\rangle$ is really the ground state, the energy difference between these two states, $\Delta E_{G,T} \equiv \langle T | H | T \rangle - \langle G | H | G \rangle$, must be positive. Evaluating this quantity, we obtain
\begin{equation} 
\begin{split}
\Delta E_{G,T}
&= \langle G | \sum_\sigma \int d \bm{r} 
\psi_\sigma^\dagger (\bm{r}) \frac{n\hat{\ell}_z}{mr^2} \psi_\sigma (\bm{r}) | G \rangle \\
&+ \langle G | \sum_\sigma \int d \bm{r} 
\psi_\sigma^\dagger (\bm{r}) \frac{n^2}{2mr^2} \psi_\sigma (\bm{r}) | G \rangle \\
\end{split} 
\label{eq.a3}
\end{equation}
where ${\hat{\ell}}_z=-i\partial_\theta$ is the angular momentum operator. When the radius $R$ of the ring is much larger than the ring width $w$, we may evaluate Eq. (\ref{eq.a3}) as
\begin{equation} 
\begin{split}
\Delta E_{G,T} = \frac{N}{2mR^2} ~n \left( n + \frac{2\ev{L_z}}{N}\right).
\end{split} 
\label{eq.a4}
\end{equation}
Here, $N$ is the total number of atoms, which is proportional to $R$ for a fixed particle density. $\ev{L_z} \simeq mRJ_T$ describes the total angular momentum of the ground state $|G\rangle$, where $J_T$ is the total current. 
\par
In the case of $J_T=O(R)$, when $R$ is sufficiently large, we can ignore the first term in the parenthesis of Eq. (\ref{eq.a4}). In this case, we can always make $\Delta E_{G,T}$ in Eq. (\ref{eq.a4}) negative by choosing an appropriate value of $n$, which, however, contradicts the assumption that $|G\rangle$ is the ground state. To avoid this contradiction, the ground state must be in the zero total angular momentum state.
\par
The two terms in the parenthesis of Eq. (\ref{eq.a4}) becomes comparable to each other when $J_T=O(R^0)$. In this case, in order to guarantee $\Delta E_{G,T}>0$, Eq. (\ref{eq.LL}) must be satisfied.



\begin{thebibliography}{99}
\bibitem{Soda} H. Shiba and T. Soda, Prog. Theor. Phys. {\bf 41}, 25 (1969).
\bibitem{Buzdin1}  For a review, see A. I. Buzdin, Rev. Mod. Phys. \textbf{77}, 935 (2005).
\bibitem{Bulaevskii} L. N. Bulaevskii, V. V. Kuzii, and A. A. Sobyanin, JETP Lett. \textbf{25}, 290 (1977).
\bibitem{Buzdin2}  A. I. Buzdin, L. N. Bulaevskii, and S. V. Panyukov, JETP Lett. \textbf{35}, 178 (1982).
\bibitem{Kanegae} Y. Kanegae and Y. Ohashi, J. Phys. Soc. Jpn. \textbf{70}, 2124 (2001). 
\bibitem{Ryazanov} V. V. Ryazanov, V. A. Oboznov, A. Yu. Rusanov, A. V. Veretennikov, A. A. Golubov, and J. Aarts, Phys. Rev. Lett. \textbf{86}, 2427 (2001).
\bibitem{Oboznov} V. A. Oboznov, V. V. Bol'ginov, A. K. Feofanov, V. V. Ryazanov, and A. I. Buzdin, Phys. Rev. Lett. \textbf{96}, 197003 (2006).
\bibitem{note00} The superfluid state with this sign change of the order parameter is sometimes referred to as the $\pi$-phase in the literature. In this paper, however, we call this superfluid state the $\pi$-junction state, in order to emphasize that the $\pi$-junction is also realized at the same time. When the sign change of the order parameter is absent, we call this state the $0$-junction state. As shown in this paper, when the $\pi$-junction state is realized in a ring trap, it induces a finite spontaneous current. To distinguish between the simple $\pi$-junction state and that with a spontaneous superflow, in this paper, we call the latter case the spontaneous current state or $\pi$-ring state.
\bibitem{FF} P. Fulde and R. A. Ferrel, Phys. Rev. \textbf{135}, A550 (1964).
\bibitem{LO} A. I. Larkin and Y. N. Ovchinnikov, Sov. Phys. JETP \textbf{20}, 762 (1965).  
\bibitem{Takada} S. Takada and T. Izuyama. Prog. Theor. Phys. \textbf{41}, 635 (1969).
\bibitem{Tsuei2} H. Hilgenkamp, Ariando, H.-J. H. Smilde, D. H. A. Blank, G. Rijnders, H. Rogella, J. R. Kirtley, and C. C. Tsuei, Nature, {\bf 422}, 50 (2003).
\bibitem{Sigrist} M. Sigrist, and T. M. Rice, J. Phys. Soc. Jpn. {\bf 61}, 4283 (1992).
\bibitem{Tsuei} C. C. Tsuei, and J. R. Kirtley, Rev. Mod. Phys. {\bf 72}, 969 (2000).
\bibitem{Harlingen} D. J. van Harlingen, Rev. Mod. Phys. {\bf 67}, 515 (1995).
\bibitem{Kato} M. Ako, M. Machida, T. Koyama, T. Ishida, and M. Kato, Physica C, {\bf 426}, 122 (2005).
\bibitem{Kulic} M. L. Kuli\'c, Phys. Rev. A \textbf{76}, 053625 (2007). 
\bibitem{Zwierlein1} M. W. Zwierlein, A. Schirotzek, C. H. Schunck, and W. Ketterle, Science \textbf{311}, 492 (2006).
\bibitem{Partridge} G. B. Partridge, W. Li, R. I. Kamar, Y.-A. Liao, and R. G. Hulet, Science \textbf{311}, 503 (2006).
\bibitem{Kashimura} T. Kashimura, S. Tsuchiya, and Y. Ohashi, Phys. Rev. A \textbf{82}, 033617 (2010).
\bibitem{Ryu} C. Ryu, M. F. Andersen, P. Clad\'e, V. Natarajan, K. Helmerson, and W. D. Phillips, Phys. Rev. Lett. {\bf 99}, 260401 (2007).
\bibitem{Mahan} G. D. Mahan, {\it Many-Particle Physics} (Plenum Press, NY, 1993) Chap. 1. 
\bibitem{Bohm} D. Bohm, Phys. Rev. \textbf{75}, 502 (1949).
\bibitem{Ohashi1} Y. Ohashi and T. Momoi, J. Phys. Soc. Jpn. \textbf{65}, 3254 (1996).
\bibitem{note10} The Bohm's theorem is an extension of the Bloch's theorem \cite{Bohm,Ohashi1}, stating the absence of finite total momentum in any thermodynamically stable states (as far as there is no external magnetic field) to the ring system.
\bibitem{note11} In the presence of the gradient $\partial_x\theta(x)$ of the order parameter phase $\theta(x)$, the total current is given by $J_T\sim N\partial_x\theta(x)$, where $N=2\pi R\rho$ is the total number of particles (where $\rho$ is the particle density). Noting that $\partial_x\theta(x)\simeq \pi/L_x\propto R^{-1}$, we find $J_T=O(R^0)$.
\bibitem{note12} Since an excess $\uparrow$-spin atom is localized to form the ferromagnetic junction, we have evaluated $N$ as $N=(N_\uparrow-1)+N_\downarrow=40+40$.
\bibitem{Shin} Y. Shin, C. H. Schunck, A. Schirotzek, and W. Ketterle, Nature (London) \textbf{451}, 689 (2008).
\end{thebibliography}
\end{document}